\documentclass{article}

\usepackage{arxiv}

\usepackage[utf8]{inputenc} 
\usepackage[T1]{fontenc}    
\usepackage{hyperref}       
\usepackage{url}            
\usepackage{booktabs}       
\usepackage{amsfonts}       
\usepackage{nicefrac}       
\usepackage{microtype}      
\usepackage{lipsum}		
\usepackage{graphicx}
\usepackage{natbib}
\usepackage{doi}

\usepackage[normalem ]{ulem}

\title{How Researchers Could Obtain Quick and Cheap User Feedback on their Algorithms Without Having to Operate their Own Recommender System}

\date{December 14, 2022}	

\author{ \href{https://orcid.org/0000-0002-8351-2823}{\includegraphics[scale=0.06]{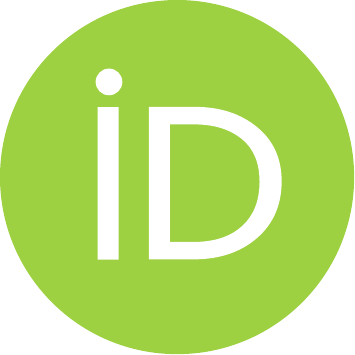}\hspace{1mm}Tobias Eichinger}\\
	Technische Universit\"{a}t Berlin\\
	Service-centric Networking\\
	Stra{\ss}e des 17. Juni,\\ 10623 Berlin, Germany \\
	\texttt{tobias.eichinger@tu-berlin.de} \\
	\And
	Ananta Lamichhane\\
	Technische Universit\"{a}t Berlin\\
	Service-centric Networking\\
	Stra{\ss}e des 17. Juni,\\ 10623 Berlin, Germany \\
	\texttt{ananta.lamichhane@campus.tu-berlin.de} \\
}

\renewcommand{\undertitle}{Technical Report}
\renewcommand{\shorttitle}{\textit{arXiv} Template}

\hypersetup{
pdftitle={A template for the arxiv style},
pdfsubject={q-bio.NC, q-bio.QM},
pdfauthor={Tobias Eichinger, Ananta Lamichhane},
pdfkeywords={recommender systems, recommender algorithms, online evaluation, offline evaluation, user study, offline user study, hybrid evaluation, user feedback collection},
}

\begin{document}
	
\maketitle

\begin{abstract}
	The majority of recommendation algorithms are evaluated on the basis of historic benchmark datasets. Evaluation on historic benchmark datasets is quick and cheap to conduct, yet excludes the viewpoint of users who actually consume recommendations. User feedback is seldom collected, since it requires access to an operational recommender system. Establishing and maintaining an operational recommender system imposes a timely and financial burden that a majority of researchers cannot shoulder. We aim to reduce this burden in order to promote widespread user-centric evaluations of recommendation algorithms, in particular for novice researchers in the field. We present work in progress on an evaluation tool that implements a novel paradigm that enables user-centric evaluations of recommendation algorithms without access to an operational recommender system. Finally, we sketch the experiments we plan to conduct with the help of the evaluation tool. 
\end{abstract}


\section{Introduction}
\label{sec:introduction}

Scholarly work on recommender systems focuses on studying recommender algorithms as part of a recommender system rather than studying entire recommender systems themselves. The focus on algorithms comes with the benefit that algorithms need not be evaluated on operational  recommender systems. Instead, it suffices to generalize algorithm inputs in order to establish generalizability of the algorithm outputs. Benchmark datasets such as the \emph{MovieLens} dataset \cite{Harper2015} or the \emph{Amazon} dataset \cite{Ni2019} represent standard examples of such generalized algorithm inputs.

The evaluation of recommender systems on the basis of historic benchmark datasets is commonly denoted offline evaluation (see for instance  \cite{Aggarwal2016recommender,Gunawardana2015,Herlocker2004}). It is the de facto standard of evaluation in the literature \cite{Jannach2012}. Offline evaluation is quick and cheap to conduct, since it neither requires operating a recommender system nor does it require to conduct a user experiment. Despite these two benefits, offline evaluation is inherently limited to objective performance measurements. It traditionally excludes the collection of subjective user feedback.

Many researchers lament that the widespread limitation to offline evaluation does not sufficiently establish the superiority of novel algorithms \cite{zanitti2018user-centric_diversity,kaminskas2016diversity_user_centric,Beel2015comparison,pu2011user_centric_resque,ge2010beyond,jones2007user_centric_metrics,mcnee2006being,ziegler2005improving_topic_diversification}. However, no prior work has addressed the problem of lowering the timely and financial burden of collecting user feedback to promote widespread user-centric evaluations. In the paper at hand, we address this problem. We aim to provide researchers, in particular novice researchers, a quick and cheap way to collect user feedback on their recommender algorithms without having to operate their own recommender system. In brief, we present the following work in progress: 
\begin{itemize}
	\item We propose a novel evaluation paradigm that may allow researchers to collect user feedback on their algorithms without having to operate their own recommender system. 
	\item We present an evaluation tool that implements the outlined evaluation paradigm. 
\end{itemize} 
Before we tend to the conceptual groundwork and the tool implementation, we first review some literature on standard evaluation methods in the field of recommender systems.

\begin{figure*}
	\centering
	\includegraphics[width=\linewidth]{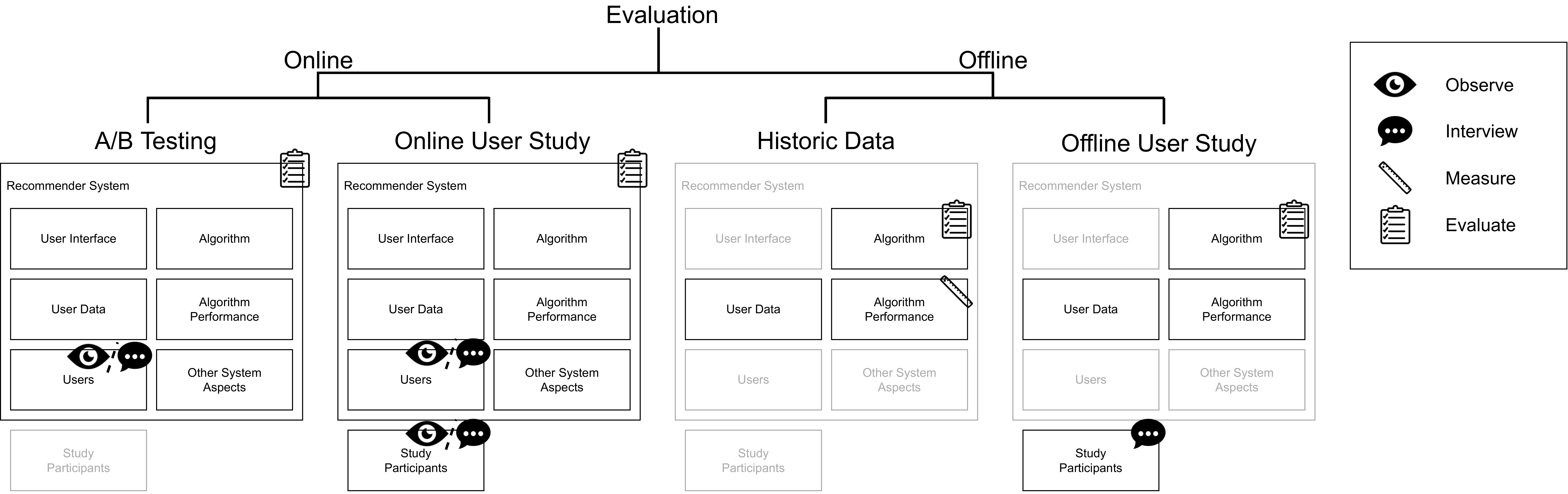}
	\caption{Comparison of standard evaluation methods. 
		Observe how online user studies are conducted on operational recommender systems (see for instance \cite{Beel2013}), while offline user studies are conducted independently from operational recommender systems 	(see for instance \cite{rossetti2016contrasting}). Researchers who do not operate their own recommender system are limited to the collection of user feedback through offline user studies. 
	}\label{fig:evaluation-taxonomy}
\end{figure*}

\section{Related Work}
\label{sec:related_work}

The literature distinguishes three standard classes of evaluation methods:  online evaluation, offline evaluation, and user studies  \cite{Aggarwal2016recommender,Gunawardana2015,Herlocker2004}. We first delineate offline and online evaluation for which we propose a novel characterization. We then tend to user studies. An overview of our proposed characterization is given in Figure \ref{fig:evaluation-taxonomy}.

\subsection{Online Evaluation versus Offline Evaluation}
\label{sec:onlineoffline}

Offline evaluation is commonly understood as evaluating algorithm performance on the basis of historic datasets. It is the de facto standard evaluation paradigm in the field of recommender systems. Any evaluation that is not based on a historic benchmark dataset is typically considered online evaluation. We propose a different characterization based on the assumption that if researchers actually had access to operational recommender systems, they would not limit their evaluation to offline evaluation on historic benchmark datasets.

We characterize \emph{online evaluation} methods as evaluation methods that are conducted on operational recommender systems, and \emph{offline evaluation} methods as evaluation methods that are conducted independently from any operational recommender system. In this view, online evaluation methods evaluate recommender systems as a whole, or rather the experience they provide to their users \cite{gomez2015netflix}, while offline evaluation methods evaluate recommendation algorithms.  Considering that algorithms do not comprise entire recommender systems, it is not surprising that online and offline evaluation methods yield disparate results \cite{rossetti2016contrasting,Beel2015comparison,ekstrand2014user,garcin2014offline,cremonesi2013user,dooms2011user}.

\subsection{User Studies}
\label{sec:userstudy}

User studies cannot be exclusively associated with online or offline evaluation. We consider a user study an online evaluation method when it is conducted on an operational recommender system, and an offline evaluation method when it is conducted independently from an operational recommender system as for instance in a laboratory environment. 

Recommender system operators have the benefit that they can recruit users as study participants. Since users have a history of interactions with the system, that is the user data that serves as input to recommender algorithms, it is possible to derive study participant recommendations in this case. However, for the majority of researchers who do not operate their own recommender system, it is not possible to derive study participant  recommendations, since no user data is available to them. 

Prior work that addresses this problem has studied questionnaire designs that yield reponses that allow to gain a good understanding of a study participant's item preferences  \cite{kostric2021,elahi2014active,pu2011user_centric_resque,rashid2008learning,rashid2002getting}. We follow this line of thought and outline the concept of mapping questionnaire responses onto  user data in the following section. The idea is that   user recommendations could serve as proxies for study participant recommendations if mapping preserves preferences.

\section{Concept}
\label{sec:concept}

We present a conceptual model for recommender systems evaluation. On the basis of this conceptual model, we describe how we propose that researchers can collect user feedback without having to operate a recommender system.

\subsection{A Conceptual Model for Recommender Systems Evaluation}
\label{sec:model}

\begin{figure}[h]
	\centering
	\includegraphics[width=\linewidth]{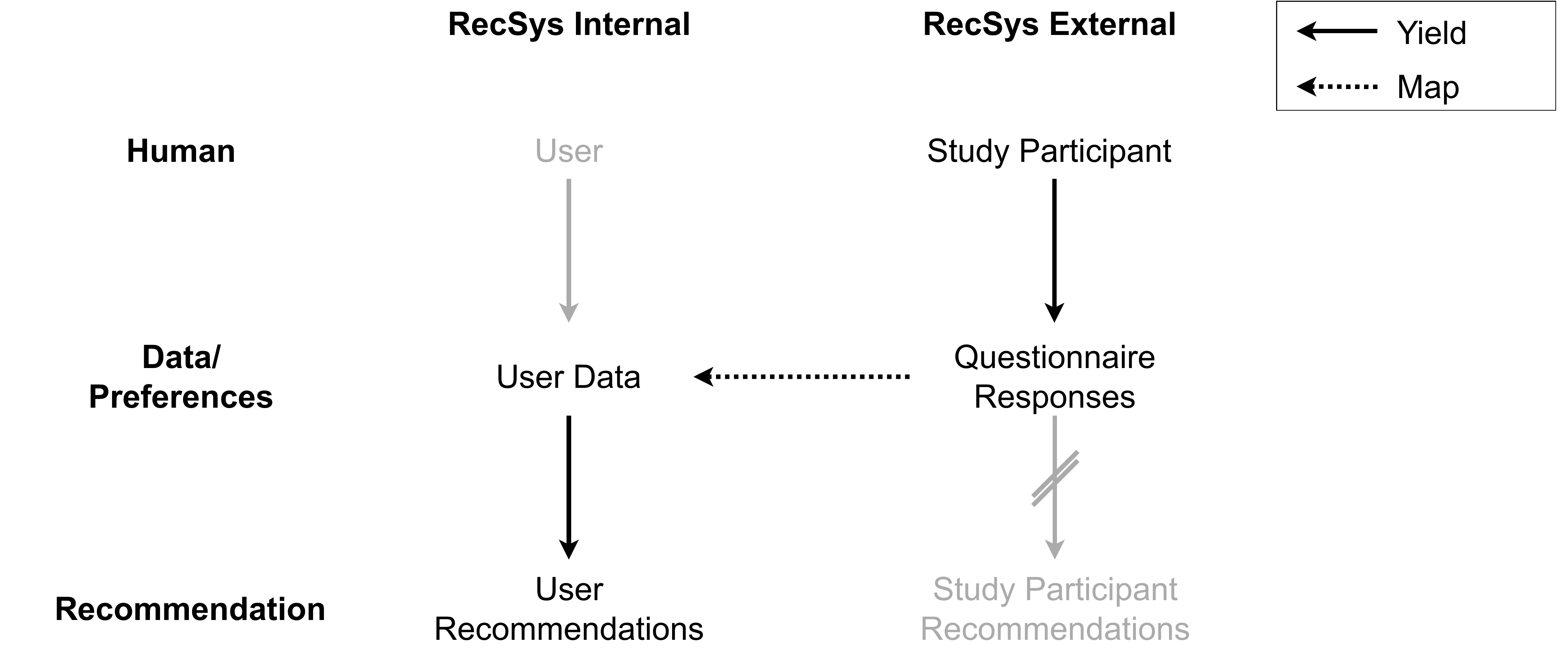}
	\caption{Conceptual model for recommender system evaluations. The model features three layers (top to bottom) and distinguishes whether an operational recommender system is given (left) or not (right). Researchers that do not have access to an operational recommender system can only use the components drawn in black for evaluation. }\label{fig:user-association}
\end{figure}

Recommender systems are particular information retrieval systems. Users in recommender systems do not formulate explicit queries. Instead, latent user preferences that are never pronounced explicitly by users act as queries. Preferences are extracted and represented by behavioral data that users produce while interacting with for instance an e-commerce shop. Such user data includes clicks, stay times, and ratings. Recommender algorithms take user data, that represent user preferences, as input and generate recommendations as output. This is depicated on the left side of Figure \ref{fig:user-association}.

A majority of recommender systems researchers does not have access to an operational recommender system. In particular, they do not have access to the users of a system. Collecting user feedback is therefore limited to user feedback from study participants of a user study who, in contrast to users, do not actively use a recommender system. Since questionnaire responses do not qualify as user data, they cannot serve as input to recommendation algorithms. This is depicated on the right side of Figure \ref{fig:user-association}.

\subsection{Mapping Questionnaire Responses onto User Data}
\label{sec:association}

User-centric evaluations of recommendation algorithms are technically infeasible when researchers cannot access an operational recommender system. On the one hand, researchers commonly have access to user data in the form of anynomized benchmark datasets, yet cannot reach out to the users whose preferences they represent (top left in Figure \ref{fig:user-association}). On the other hand, researchers can recruit study participants and collect questionnaire responses yet cannot derive recommendations, since questionnaire responses do not represent user data (bottom right in Figure \ref{fig:user-association}). We propose the following idea to resolve this asymmetry.

We propose to identify questionnaire responses that elucidate the preferences of a study participant with  user data that elucidate the preferences of a user in a benchmark dataset (see the dotted arrow in Figure \ref{fig:user-association}). More specifically, researchers collect questionnaire responses from a study participant and map them to the user data of a user in a publicly available benchmark dataset. If the mapping preserves the preferences of study participants and users, we argue that we can view study participants as proxies to users. 
Also, we argue that we can view user recommendations as proxies to study participant recommendations.

\section{Implementation}
\label{sec:implementation}

We present the implementation of our evaluation tool. We then sketch the experiments we plan to conduct with the help of the evaluation tool.

\subsection{Tool Implementation}
\label{sec:testbed}

The evaluation tool is implemented as a web application written in the \emph{Python3} and \emph{JavaScript} programing languages.\footnote{\url{https://github.com/ananta-lamichhane/surveyapp}} The web application allows researchers to conduct user studies to collect subjective feedback on the recommendations their algorithms produce. We present the tool's functionality by describing the distinct researcher and study participant workflows. 

\textbf{Researcher Workflow:} Researchers create a user study by providing (a) a study title and description, (b) a benchmark dataset, (c) a mapping mechanism that maps questionnaire responses to user data, (d) the type of questions in a questionnaire, (e) two sets of pre-calculated user recommendations for all users in the benchmark dataset, (f) user-centric evaluation metrics with respect to which the pre-calculated user recommendations are to be compared, and finally (g) the email addresses of the survey participants. After creation, starting a user study sends invitations to all study participants. Results are  downloaded after the study has been closed. 

\begin{figure}[t]
	\centering
	\includegraphics[width=\linewidth]{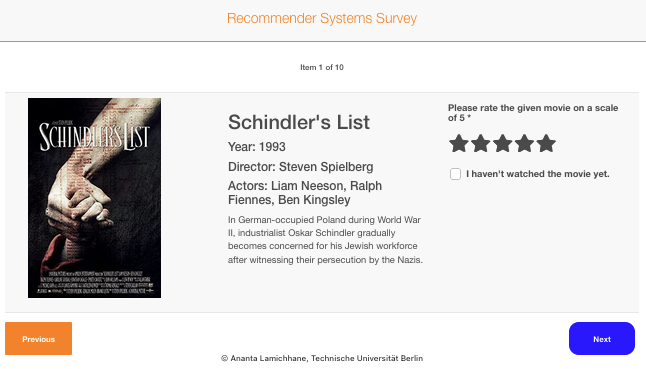}
	\caption{Screenshot of a question in a sample questionnaire to be filled in by a study participant.}\label{fig:question}
\end{figure}

\textbf{Study Participant Workflow:} Invited study participants answer an initial set of questions whose responses are mapped to user data in the benchmark dataset. A screenshot of an initial question is depicted in Figure \ref{fig:question}. After mapping the initial question responses to user data, study participants are prompted with two sets of user recommendations that have been pre-calculated for the mapped user data (see (e) in the researcher workflow). Study participants answer a final set of questions that compare the prompted user recommendations with respect to user-centric evaluation metrics such as novelty, diversity, or serendipity.

\subsection{Experiments to Conduct with the Tool}

The novel evaluation paradigm is only meaningful if it is possible to map questionnaire responses onto user data such that the underlying preferences between study participants and users are preserved. The effectiveness with which preferences are preserved can be measured on all three layers of the conceptual evaluation model shown in Figure \ref{fig:user-association}. 

\textbf{Human-Layer Experiments:} We recruit study participants and conduct a user study. We map the questionnaire responses on the initial set of questions onto user data. After mapping, we do not show the final set of questions on recommendations, yet instead prompt the mapped user data itself to the study participant and ask to what degree the prompted user data agrees with the study participant's preferences. The higher the degree of agreement is, the better the mapping preserves preferences between study participants and users.

\textbf{Data-Layer Experiments:} We numerically measure the similarity between questionnaire responses and the user data they are mapped onto. The type of similarity measure that can be used depends on the type of questionnaire responses received in the initial set of questions. Analogously to human-centric experiments, the higher the degree of similarity is, the better the mapping preserves the preferences between study participants and users.   

\textbf{Recommendation-Layer Experiments:} We select a subset of users in a benchmark dataset and treat them as human study participants. We then simulate a user study by filling in questionnaires on the basis of the selected users' user data. After simulation, we measure the percentage of simulated questionnaire responses that have been mapped correctly. The higher the percentage is, the better the mapping preserves the preferences between study participants and users.

\section{Summary and Lookout}
\label{sec:conclusion}

We present an evaluation tool that aims to allow researchers to collect user feedback on their  recommender algorithms without access to an operational recommender system. The tool addresses  the lack of user-centric evaluations in the recommender systems literature. It lowers the timely and financial burden imposed by user-centric evaluations in the sense that researchers do not need to establish and maintain their own operational recommender system to conduct research. 

The evaluation tool is based on a novel evaluation paradigm that leverages benchmark datasets in conjunction with traditional user studies. Applying the novel evaluation paradigm with the presented tool is only meaningful if it is possible map the questionnaire reponses of a study participant onto the user data of a user in a benchmark dataset such that both have similar preferences. We outline a set of experiments we will conduct on the presented evaluation tool that will or will not allow to substantiate the soundness of the proposed evaluation paradigm. 

Hopefully, we will be able to show the soundness of the proposed evaluation paradigm in the upcoming year. If so, the evaluation tool will enable the majority of recommender systems researchers that do not have access to an operational recommender system to obtain quick and cheap feedback on their algorithms. Then the field might open up widely towards user-centric evaluations.

\bibliographystyle{unsrtnat}
\bibliography{main}  

\end{document}